\documentclass[12pt]{iopart}

\input epsf.sty

\begin{document}
\title[Algorithm of a clinical beam-angle optimization system]{Algorithm and performance of 
a clinical IMRT beam-angle optimization system}


\author{David Djajaputra$^1$\footnote[1]{Present address: Department of Radiation Oncology 
and Molecular Radiation Sciences, Johns Hopkins University, 401 N. Broadway, Baltimore, MD 21231, USA} 
Qiuwen Wu$^1$, Yan Wu$^1$, and Radhe Mohan$^2$}

\address{$^1$\ Department of Radiation Oncology, Virginia Commonwealth University 
Health System, Box 980058, Richmond, VA 23298, USA}

\address{$^2$\ Department of Radiation Physics, Unit 94, The University of Texas,
M D Anderson Cancer Center, Houston, TX 77030, USA}

\begin{abstract}
%
%
This paper describes the algorithm and examines the performance of an
IMRT beam-angle optimization (BAO) system. 
%
%
In this algorithm 
successive sets of beam angles are selected from a set of predefined
directions using a fast simulated annealing (FSA) algorithm. An IMRT 
beam-profile optimization is performed on each generated set of beams. 
The IMRT optimization is accelerated by using a fast dose calculation
method that utilizes a precomputed dose kernel. A compact kernel is
constructed for each of the predefined beams prior to starting the
FSA algorithm. The IMRT optimizations during the BAO are then 
performed using these kernels in a fast dose calculation engine. 
This technique allows the IMRT optimization to be performed more
than two orders of magnitude faster than a similar optimization
that uses a convolution dose calculation engine. Any type of 
optimization criteria present in the IMRT system can be used in 
this BAO system. An objective function based on clinically-relevant
dose-volume (DV) criteria is used in this study. This facilitates 
the comparison between a BAO plan and the corresponding plan 
produced by a planner since the latter is usually optimized 
using a DV-based objective function. 
%
%
A simple prostate case and a complex head-and-neck (HN) case were 
used to evaluate the usefulness and performance of this BAO method. 
For the prostate case we compared the BAO results for 3, 5, 
and 7 coplanar beams with that of the same number of equispaced 
coplanar beams. For the HN case we compare the BAO results for 7 and 9
{\it non}-coplanar beams with that of 9 equispaced coplanar
beams. In each case the BAO algorithm was allowed to search 
up to 1000 different sets of beams. The BAO for the prostate
cases were finished in about 1--2 hours on a moderate 400 MHz 
workstation while that for the head-and-neck cases were 
completed in 13--17 hours on a 750 MHz machine. No a priori
beam-selection criteria have been used in achieving this performance.  
%
%
In both the prostate and the head-and-neck cases, 
BAO is shown to provide improvements in plan
quality over that of the equispaced beams. The use of DV-based
objective function also allows us to study the dependence of the 
improvement of plan quality offered by BAO on the DV criteria 
used in the optimization. We found that BAO is especially useful
for cases that require strong DV criteria. 
%
%
The main advantages of this BAO system are its speed and its 
direct link to a clinical IMRT system.

\end{abstract}


\ead{djajaputra@jhmi.edu, qwu@vcu.edu}
\maketitle

\noindent {\bf List of abbreviations}

\begin{tabbing}
BAO    \qquad \= Beam Angle Optimization \\
CRT    \> Conformal Radiotherapy \\
CSA    \> Classical Simulated Annealing \\
DV     \> Dose Volume \\
DVH    \> Dose Volume Histogram \\
EUD    \> Equivalent Uniform Dose \\
FSA    \> Fast Simulated Annealing \\
HN     \> Head and Neck \\
IMRT   \> Intensity Modulated Radiation Therapy \\
IO-SC  \> Intensity Optimization with Superposition Convolution \\
IO-TLP \> Intensity Optimization with Table Lookup \\
OAR    \> Organ at Risk \\
OF     \> Objective Function \\
SA     \> Simulated Annealing \\
SC     \> Superposition Convolution \\ 
TLP    \> Table Lookup \\
TPS    \> Treatment Planning System \\
VOI    \> Volume of Interest \\
\end{tabbing}

\section{Introduction}

The beam-angle optimization (BAO) in intensity-modulated 
radiation therapy (IMRT) aims to find, for each patient-specific case,
a set of beam angles that can produce a three-dimensional (3D) dose 
distribution that conforms most closely to the prescribed one.
In general, selecting the best set of beam angles that can achieve 
this goal is a highly non-intuitive process. This is already true in
3D conformal radiation therapy (CRT), where each beam has a preset profile 
inside its beam's-eye-view aperture and only the beam weights 
are optimized, but especially more so
in IMRT. The optimal distribution of the total dose in IMRT is the 
result of a complex interplay among a large number of beamlet intensities 
from multiple beam angles and therefore there 
is no a priori way of predicting the dependence 
of the objective function (OF) on the beam angles. The optimizer has
to balance the requirement of creating a prescribed dose distribution 
at the target with that of giving as little dose as possible to
the surrounding normal tissue and organs at risk (OARs). 
Stein \etal (1997) provided a well-known counter-intuitive example of the 
necessary compromise that often has to be struck to achieve 
this balance between target coverage and OAR sparing: sometimes 
the beams have to go directly through an OAR to achieve an 
optimal distribution.

The complex dependence of the dose distribution on the beam 
angles and their beamlet intensities gives rise to a multiple
local minima problem. The fact that the OF has
a highly non-convex dependence on its variables has so far precluded
any deterministic approach to its minimization and necessitated
a brute-force assail to this problem, namely a global search method.
This approach requires one to perform an IMRT optimization on each one of
a large number of possible combinations of beam angles.
Two main hurdles have to be overcome, however, in making this 
highly-iterative approach practical and can be used routinely in a clinical setting.
The first hurdle is the substantial amount of time that is required 
to perform an IMRT beam-profile optimization for a single set of beam 
angles. Depending on the complexity of the case and the algorithm
used in the optimization, the required time already ranges from several 
minutes to a few hours on a typical workstation. The second hurdle
is the huge size of search space involved in the beam-angle search.
To rigorously search for, e.g., the best 5 gantry angles out of 180 possible 
directions would require one to search over $C^5_{180} (\sim 10^9)$ 
possible combinations. A BAO system that is based on an exhaustive search
would therefore, most likely, be too slow for utilization in a clinical setting.

The recent past has seen a limited but steady flow of publications 
on this topic. A few papers have provided theoretical analysis of 
the BAO problem (Bortfeld and Schlegel 1993, Jackson \etal 1994, 
Braunstein and Levine 2000).
Bortfeld and Schlegel (1993) considered the BAO for the 3D CRT, 
as opposed to IMRT, and concluded that, for dose-based criteria, 
the optimum beam configuration for multiple-beam irradiation tends to be an even 
distribution over the $2\pi$ angular range for the gantry. 


Rowbottom \etal (1998, 1999a, 1999b) have proposed several approaches
for BAO as applied to the 3D CRT. Dose-based OF was used 
in the optimization of the beam weights and simulated annealing 
algorithm was used to select the beam orientations in a reduced
search space of coplanar (Rowbottom \etal 1998) and non-coplanar
(Rowbottom \etal 1999a) beams. Some a priori criteria were also used 
in the beam-orientation selection: the beam orientations were 
constrained to be separated by at least $35^\circ$ and 
parallel-opposed beams were disallowed. In addition, a single-beam
cost function (Oldham 1998), based largely on the geometry of the 
target, was used to rank the beam orientations
in the search space according to their likelihood for being a member
of the optimal set and only the top 10\% of the set were searched 
in the BAO. Rowbottom \etal (1999b) also attempted a more intelligent 
search using an artificial neural network technique. The results of this 
search, however, seem to be inferior to those that were obtained using their 
earlier simulated annealing method. 

Pugachev and Xing recently proposed several approaches to BAO for the IMRT
(Pugachev \etal 2000, 2001, Pugachev and Xing 2001, 2002). In their
earliest proposal, a non-iterative filtered back projection (FBP) method, using 
simple exponential attenuation for parallel X-ray beams in their dose
calculation, was used to solve the inverse problem in IMRT while the beam orientations 
were optimized using a simulated annealing method. This fast 
FBP approach, however, seems to preclude the use of OFs
other than a dose-based OF. In Pugachev \etal (2001), 
an iterative method (Xing and Chen 1996) was used to optimize the 
beam profiles and non-coplanar beams were used in the search space.
A typical prostate case in this study required about 100 hours to be 
optimized by searching over 5000 combinations of beam angles.
The use of the beam's-eye-view dosimetrics (BEVD) (Pugachev and Xing
2001, 2002) allowed them a more efficient sampling of the search space:
The sets of beams that have low BEVD score, i.e. those that are not
likely to produce an optimal dose distribution, are immediately rejected
without going through the beam-profile optimization. It was claimed
that the BEVD-guided sampling can improve the speed by a factor of 10
while maintaining the same capacity to locate the global minimum as 
that of an unguided sampling. 

It is interesting to note that most of previous approaches 
to BAO used the dose-based OF (Bortfeld and Schlegel
1993, Rowbottom \etal 1998, 1999a, 1999b, Pugachev 2000, 2001, 
Pugachev and Xing 2002). Das \etal (2003) proposed an IMRT beam orientation
selection method that is based on target equivalent uniform dose (EUD) 
maximization. The objective is to maximize the EUD on the target while 
simultaneously imposing DV constraints on the OARs. These two objectives 
were combined into a single OF by a simple scale factor and it would 
be interesting to compare the performance of this combined OF with 
that of a strictly DV-based or EUD-based OF. 

The approach that we take in this paper is to use a fast dose
calculation engine that significantly reduces the time required 
for the IMRT optimization of each given set of beam angles. 
This algorithm relies on the use of a compact, albeit approximate,
dose kernel (Wu \etal 2003). The main purpose of this paper is to show that 
this fast algorithm is sufficiently accurate to be used as a dose engine
in a BAO. Although simulated annealing has been used in several previous works 
on BAO, in this paper we show that the {\it combination} of simulated annealing, in 
particular the FSA algorithm, and the fast dose engine that we proposed
recently can provide a practical, i.e. fast and useful, BAO system. 
In addition, by focusing the improvement on the speed of the dose engine 
in our system, we are also able to retain all other aspects of normal, 
i.e. clinical, IMRT optimization in our BAO. Thus our BAO system allows 
a DV-based OF, or indeed any other types of OF, to be used for the IMRT 
optimization for each given set of beam angles generated during the BAO. 
The promise of a BAO system is to provide better
alternatives to plans produced by a planner, which in clinical
settings are commonly generated by a DV-based optimization. It is therefore
important to use a DV-based, instead of a dose-based, optimization 
in a clinical IMRT BAO system so that a direct comparison can be effected.
If the BAO plan is deemed by the planner to be significantly better, it
can be used for treatment; otherwise, the plan produced manually
by the planner is retained. Thus the BAO can be used to automatically  
check if the best plan of the planner can still be improved by 
beam-angle adjustment. Starting with an already acceptable plan, the BAO
is executed with the same DV constraints and parameters and its plan 
will be used if it is significantly better
and can be produced within an allowable time. 

It should be noted that no a priori assumption is made regarding the 
selection of beam orientations in this work. Although incorporation
of some criteria, such as constraints on the angular separation between
adjacent beams, is likely to improve the performance of the system,
we feel that such topics require separate careful studies. These
additional beam-selection criteria will be incorporated in further studies.

\section{Methods and materials}

In this work, we propose an efficient BAO method 
that combines a fast dose calculation engine, 
which we call the table lookup (TLP) method (Wu \etal 2003), with an effective 
global search method based on the fast simulated annealing (FSA) algorithm. 
We have recently implemented this method in a program
that interfaces with our in-house clinical IMRT 
system and studied its performance and potential. 

The following subsections will detail the components of
our BAO implementation. Section \ref{BAO} discusses the flow diagram of 
our BAO system and the algorithm that we use to select successive sets
of beams from the search pool (the FSA algorithm). The dose engine
that we use as a workhorse in the BAO (the TLP method) is presented
in section \ref{TLPSection} while section \ref{OFSection} describes
the DV OF that we use to guide the IMRT optimization.  
Details on how the TLP method is used to accelerate our IMRT 
optimization and discussion on its accuracy are 
provided in section \ref{OptimizationSection}. 
The cases that were used to test our BAO system
in this work are outlined in section \ref{MaterialsSection}.

\begin{figure}
\begin{center}
\end{center}
\vskip -0.4in
\caption{\label{Flowchart}Flow diagram of a BAO using 
the simulated annealing method. The system has been implemented 
to accept non-coplanar beams in the search pool shown 
in the topmost box. The kernel is
calculated using an accurate SC dose engine of the TPS for 
each beam in the search pool. This kernel is used during
the IMRT beam-profile optimization for dose calculation
using the TLP method. Note that in the final step, once
the optimal set of beams is selected, a final IMRT
beam-profile optimization is performed using the 
SC method before the plan is output.} 
\end{figure}

\subsection{Beam angle selection with fast simulated annealing}
\label{BAO}

Figure \ref{Flowchart} displays the flow diagram of our BAO system
which is interfaced with an in-house IMRT system.
The IMRT system, in which the planner sets general optimization 
parameters including the number and energy of beams whose directions 
are to be optimized, is in turn interfaced with a commercial treatment
planning system (TPS) Pinnacle$^3$ (Philips Oncology System, CA, USA). 
This in-house IMRT BAO program, which is external
to the TPS, communicates with the TPS through an application programming 
interface, provided by the TPS vendor, called PinnComm. This allows the 
BAO and the IMRT programs to, e.g., request the TPS to perform a dose calculation.
 
The BAO is performed using an FSA algorithm.
Prior to entering the SA loop, the planner defines a 
search pool of beams. We use 180 equispaced beams 
for a coplanar search pool and 216 beams (72 equispaced beams at couch
angle of 0$^\circ$, 45$^\circ$, and 315$^\circ$) for a non-coplanar
search in this work. For each beam in the search pool, the TPS is called
by the BAO program to calculate an accurate dose for a uniform intensity
profile. The jaws are automatically set for each beam to sufficiently
cover the target with some margins. The kernel is then extracted from the
TPS using the TLP method (section \ref{TLPSection}). 
Note that we do {\it not} optimize the number of beams in this study.

\subsubsection{Fast simulated annealing: temperature lowering schedule.}

The BAO selects the optimal set of beams from the search pool. Successive 
sets of beam angles are generated during the BAO by using the FSA 
algorithm (Szu and Hartley 1987). The efficacy of this global search
method has been proven in many fields, including IMRT and beam-angle
optimization (Webb 1989, 1991, Mageras and Mohan 1993, Xiang and Gong 2000,
Pugachev 2000). The FSA is known to be much more efficient than the 
classical SA (CSA) algorithm (Kirkpatrick \etal 1983). To converge to 
a global minimum with probability one, the CSA requires the annealing
temperature to be lowered with an inverse-logarithmic schedule: 
$T = T_0/\ln(1 + i)$, where $i$ is the iteration number 
(Geman and Geman 1984). On the other hand, the FSA allows the temperature
to be lowered with a much faster schedule (Szu and Hartley 1987): 

\begin{equation}
T = {T_0 \over 1 + i}.
\end{equation}

\noindent The SA temperature $T$ is used in determining whether a new set
of beam angles is to be accepted or rejected. Note that temperature
in this context should be understood only as an optimization 
parameter and it does not have any relation, except in analogy, to the 
temperature of any physical system. A move generated by the FSA is
accepted with a Metropolis probability (Metropolis \etal 1953):

\begin{equation}
p = {\rm min} \ \big[ 1, e^{- \Delta E / T} \big],
\label{Metropolis}
\end{equation}

\noindent where $\Delta E$ is the change of the OF (score). Thus a 
downhill move is always accepted while an uphill move is accepted 
with a probability $e^{- \Delta E / T}$. The Metropolis probability
allows the optimization to simulate the behavior of a physical system
in a thermal equilibrium while the SA 
temperature schedule controls the lowering of temperature such that 
the system is effectively at a thermal equilibrium at any time. As in
a physical system, this recipe ensures that local minima are properly 
sampled and the global minimum is reached as the temperature is 
lowered to zero.

It should be noted that $\Delta E$ in equation \ref{Metropolis} is the
difference between the {\it optimal} score (final value of objective function) 
of the current trial set of beams and that of the last accepted set of beams.
Each of these scores is the output of a {\it beam-profile} optimization 
performed on the corresponding set of beams. Details of the beam-profile 
optimization is not needed by the FSA algorithm whose task is to generate
subsequent sets of beams and accept or reject each set based on the value of
its {\it optimal} score. 
  
\subsubsection{Fast simulated annealing: sampling distribution.}

The relative efficiency of the FSA over the CSA is related to
the sampling distribution that it uses. At each BAO iteration, FSA 
samples the {\it changes} to the current beam angles ${\bf x} = 
(\theta_1, \theta_2, \cdots, \theta_n)$ from a {\it long-ranged} 
Cauchy-Lorentz distribution: 

\begin{equation}
p(\Delta {\bf x}) \propto [ (\Delta {\bf x})^2 + W(T)^2 ]^{-(n+1)/2},
\end{equation}

\noindent where $n$ is the number of beam angles (the dimension
of the vector {\bf x}). In contrast to FSA, the 
CSA uses a {\it short-ranged} Gaussian
sampling distribution: $p(\Delta {\bf x}) \propto \exp [
- (\Delta {\bf x})^2 / 2 W(T)^2 ]$, where $W(T)$ is the 
temperature-dependent width of the sampling distribution.  
Search-space regions farther away than $W(T)$
from the current position are practically not sampled by the 
CSA distribution and therefore CSA requires a larger number
of iterations than FSA to thoroughly sample the entire search space.

Several different choices for the definition of $W(T)$ are available
in the literature (Mageras and Mohan 1993) and, in general, the ``best'' choice for an 
application can only be found by trial and error. For the FSA work reported in this 
paper, we have used a simple form $W(T)^2 = \alpha T^2$. The value
of the range parameter $\alpha$ was chosen such that even at the lowest temperature ($T=1$
in this paper) each beam direction still has 1\% probability to jump to 
another direction ``farthest away'' from it (e.g., from 0$^\circ$ to 180$^\circ$
gantry angle). Thus at any temperature, the current trial set of beams always has a finite 
probability to jump to any other point in the search space. The ability to sample
regions beyond the local neighborhood is imperative in the simulated annealing method 
to avoid being trapped in a local minimum.

In generating the random move for the set of beam angles ${\bf x} = 
(\theta_1, \theta_2, \cdots, \theta_n)$ we use the approach where 
the orientations of {\it all} beams in the current set are simultaneously
updated. This has been shown (Mageras and Mohan 1993) to be more effective
than the alternative approach, in which one beam is picked randomly
from the current set and its orientation is moved by a random distance
which is picked from the sampling distribution (Das \etal 2003).

\subsection{Fast dose-calculation method}
\label{TLPSection}


The total dose at a given voxel, $D_i$, is related linearly to the beamlet 
weights $W_j$ by the dose kernel $K_{ij}$:

\begin{equation}
D_i = \sum_j K_{ij} W_j.
\label{DoseFormula}
\end{equation}

\noindent Note that in this equation the beamlet index $j$ comprises also 
the beam index; the index $j$ runs over all beamlets in all beams.
The dose kernel $K_{ij}$ is a sparse matrix. At each dose voxel, significant
contributions to its dose come only from a limited number of beamlets 
in each beam, the largest contribution being from the beamlet that directly
traverses the dose voxel. The sparse nature of the kernel matrix can 
be shown numerically by, e.g., plotting the magnitude of the elements
of the square matrix $K^tK$ (Crooks and Xing 2001). This also gives a measure
of the overlap between beamlets.

The fact that the kernel matrix is sparse has been used to devise a fast,
albeit approximate, table-lookup (TLP) dose calculation method, 
that is useful for IMRT optimization (Wu \etal 2003). This efficient method 
was proposed to improve the performance of applications that involve a 
great number of dose calculations. In the present work we have used this 
method as the fast dose calculation engine for our IMRT BAO. In the TLP 
method, each dose voxel is assigned one, and only one, beamlet from each 
beam, namely the beamlet whose raytrace intersects the voxel. The dose to
the voxel is initially calculated using an accurate method, e.g., the 
superposition/convolution (SC) method, for a uniform intensity profile 
for all beams. The TLP kernel is then calculated using the following formula:

\begin{equation}
K_{ij} = \cases{D_i/W_j, &${\rm if \ voxel \ } i \ {\rm is \ on \ the \
path \ of \ beamlet \ } j{\rm ,}$\cr 0, &${\rm otherwise.}$}
\label{KernelFormula}
\end{equation}

\noindent Once the kernel is calculated, dose calculation in subsequent 
iterations is done by looking up the kernel. Note that in addition to 
improving the speed of the calculation, the TLP method also greatly reduces the 
number of kernel elements that need to be saved and read, and therefore the 
time needed to access them. This number, 
for each beam, is equal to the number of dose voxels because of the 
one-to-one dose-beamlet correspondence (Wu \etal 2003). The use of this
fast dose-calculation method in our IMRT optimization will be further
described in section \ref{OptimizationSection}.

\subsection{Objective function}
\label{OFSection}

For each trial set of beam angles generated during the BAO, an IMRT optimization is
performed to optimize the beam profile. This IMRT optimization is implemented
by using an objective function, or score, that provides a measure 
of ``distance'' between the current dose distribution and the desired 
distribution. Starting from an initial distribution, the objective function
is iteratively minimized by the optimizer.

Several types of objective function (OF) exist in the literature and implemented
in clinical IMRT optimization systems. Among these are the dose-based, 
DV-based, and the biology-based functions. The dose-based OF is
the most straightforward conceptually and the easiest to implement (Webb 1989):
it uses a penalty which is a quadratic function of the 
difference between the actual dose and the desired dose, which is usually the 
prescription dose for the tumor and zero for the organs at risk (OARs).

The DV-based objective function is among the ones that are most useful in
IMRT optimization (Bortfeld \etal 1997, Spirou and Chui 1998, Wu 2000). 
It has the advantage of being more intuitive
than the biology-based OFs and more flexible than, since it is the superset
of, the dose-based OF. The DV-based optimization typically requires more
iterations to achieve convergence than the dose-based optimization. Furthermore,
it requires the dose-volume histogram (DVH) to be computed after each 
dose calculation which adds to the total optimization time. In general, however,
DV objective requires less adjustment of the optimization parameters to achieve 
a suitable plan (Wu 2000). Although multiple local minima exist in a DV-based
IMRT optimization (they are absent in dose-based optimization), it has been
shown (Wu and Mohan 2002, Llacer \etal 2003) 
that they are not expected to affect the outcome of 
optimization using gradient techniques in any clinically significant way.


For the purpose of establishing our notation for later reference in the rest
of this paper, here we present a brief 
description of our implementation of the DV OF. There are two types 
of constraint in a DV-based optimization, each is 
specified by a set of dose-volume values $(D_a,V_b)$. The first type of 
constraint (upper constraint) requires the volume receiving dose greater than
$D_a$, which is denoted by $V(D_a) = V_a$, to be {\it less} than $V_b$.
This constraint is implemented by searching for the dose value $D_b$ 
such that $V(D_b) = V_b$. A penalty is charged if $V_a > V_b$, which 
is equivalent to $D_a < D_b$ since the DVH is a non-increasing function
of dose. The upper constraint for a DV-based objective function 
therefore has the following form:

\begin{equation}
f_1 = p_1 \sum_i H(D_b - D_i) \cdot H(D_i - D_a) \cdot (D_i - D_a)^2,
\quad {\rm upper \ constraint}.
\end{equation}

\noindent Here $H(x)$ is a Heaviside step function defined by

\begin{equation}
H(x) = \cases{1, &$x > 0,$\cr 0, &$x \leq 0.$}
\end{equation}

\noindent An upper constraint is used to limit the dose to an OAR
or the hot-spot regions in a target.
The index $i$ in the summation runs over all voxels in the OAR, while
$p_1$ is the penalty prefactor. The constraint's objective is to reduce 
the volume of the OAR that receives dose greater than $D_a$; it therefore
penalizes voxels that have dose $D_i > D_a$. Note, however, 
that voxels that have dose greater than $D_b$ are not penalized since the 
total volume of such voxels is already lower than $V_b$.

A lower constraint is used to limit the cold-spot regions in a target.
This constraint requires $V_a$, the volume receiving dose greater than $D_a$, 
to be {\it greater} than $V_b$. The penalty is imposed if $V_a < V_b$, or 
equivalently if $D_a > D_b$:

\begin{equation}
f_2 = p_2 \sum_i H(D_a - D_i) \cdot H(D_i - D_b) \cdot (D_i -D_a)^2,
\quad {\rm lower \ constraint}.
\end{equation}

In an IMRT optimization, DV constraints are assigned by the planner to the 
volumes of interest (VOIs). Each VOI can have more than one constraint;
these constraints control the shape of the DVH curve for the corresponding
VOI. Since there is no specific desired dose for each VOI in a DV-based
optimization, in contrast to dose-based optimization, the quality of the 
DVH of one or more VOIs can be sacrificed slightly by the optimizer to 
improve those of some other VOIs. The importance of each DV constraint,
relative to the other existing constraints, is set by the relative value of 
its penalty prefactor.

\subsection{Intensity optimization for each set of beam angles}

\label{OptimizationSection}


For each trial set of beam angles in the BAO, an iterative IMRT 
optimization using the Newton's method is performed to optimize the beam profiles
by minimizing a
DV-based objective function. At each iteration during the IMRT optimization 
each beamlet weight $W_j$ is updated to $W_j + \delta W_j$. The size 
of this update $\delta W_j$ depends on the current values of dose and the kernel: 
$\delta W_j = F \big[ \{ D_i \}, \{ K_{ij} \} \big].$
The procedure for computing $\delta W_j$ can be found in
Wu and Mohan (2000) and will not be repeated here. 
Our IMRT optimization program is interfaced with a commercial
3D TPS. For a given set of beams, we start from a 
uniform intensity profile for all beams and compute an accurate dose 
$\{D_i\}$ using the dose calculation engine (which is based on the
SC method) of the TPS. Kernel $\{K_{ij}\}$ is then calculated for 
the iterative optimization using the TLP method (equation \ref{KernelFormula}). 

\subsubsection{Intensity Optimization with Superposition Convolution (IO-SC) Method.}

Note that dose and kernel enter independently in the 
computation of the beamlet weight update $\delta W_j$. 
This allows different levels of approximation to be used for dose and the kernel.
Ideally both dose and kernel are computed accurately using, e.g., an SC
method. However, for an IMRT system that is interfaced with a commercial 
TPS, the kernel is usually not directly available from the TPS, although the dose is.
In the IO-SC method (Wu and Mohan 2000), $\delta W_j$ is obtained using an accurate 
dose and an approximate kernel at each iteration. The accurate dose, 
that incorporates scatter effects, is obtained from 
the SC dose engine of the TPS. The approximate kernel, on the other hand, is obtained 
from the accurate SC dose using equation \ref{KernelFormula}. Although scatter effects
are neglected in this approximate kernel, they are partially accounted for in the 
calculation of $\delta W_j$ since a full-scatter dose is used. The TLP method of 
constructing an approximate kernel has been found to be useful in 
practice (Wu \etal 2003) and this IO-SC method has been used for IMRT routinely
in our clinic. The main limitation of the IO-SC 
method is the fact that the dose has to be computed at each iteration
using the SC dose engine from the TPS. The total optimization time is 
then strongly dominated by the lengthy time spent on repeated dose 
calculations.

\subsubsection{Intensity Optimization with Table Lookup (IO-TLP) Method.}

For the purposes of BAO, we introduce a further approximation
to the computation of $\delta W_j$ which provides a significant speed up in 
the calculation time. Two different methods of dose calculation
are used in the IO-TLP method. An {\it initial} dose is calculated accurately 
using the SC method of the TPS. A kernel is then generated from this accurate
dose using equation \ref{KernelFormula}. In {\it subsequent} iterations, instead 
of computing a fully accurate dose using the SC method, dose is calculated by 
looking up the kernel, i.e., using equation \ref{DoseFormula}. Thus {\it both} dose
and kernel are approximate in the IO-TLP method as opposed to only the kernel being
approximate in the IO-SC method. Note that 
scatter effects are not totally absent in the IO-TLP method because a 
full-scatter dose is always used as the initial dose for the beam-profile
optimization.

The TLP method provides a substantial speed advantage over the SC method.
Since it is based mostly on memory lookup operation, each TLP dose
calculation, which is the dose engine in the IO-TLP method, is typically
at least two orders of magnitude faster than the corresponding SC
calculation, which is used in the IO-SC method. In comparing the speed of 
these two optimization methods, we need to consider two separate
situations: (1) a single IMRT optimization for a given set of beams;
and (2) the BAO in which a large number of sets of beams (1000 in
this work) are optimized consecutively. In the single IMRT case,
the IO-TLP method uses the TLP dose engine to calculate the total
dose at each iteration {\it except} the first one where SC is used
to compute the kernel. The IO-SC method, on the other hand, uses the 
SC dose engine at each iteration. The speed ratio between the IO-TLP
and IO-SC methods in this case is given roughly by the total number
of iterations needed to converge the optimization, which is usually
about 10--20 in our IMRT system. In the case of BAO, the kernel 
for each beam in the search pool is computed,
using the SC method, prior to any IMRT optimization. During
the beam-profile optimization of each given set of beams, {\it no}
SC dose calculation is needed because the kernels have all been 
computed and can be quickly read from memory by the TLP method.
Except for the time overhead spent in the initial calculation
of the kernels, which is small relative to the total BAO time
(table \ref{CPUTime}) the speed ratio in this case is given 
directly by the speed ratio between the TLP and the SC methods.
Thus our BAO system is at least two orders of magnitude 
faster than it would be if we used the SC method as the dose
engine for beam-profile optimization.


\begin{figure}
\begin{center}
\end{center}
\vskip -0.4in
\caption{\label{LookupPinnacle}Good correlation exists between the optimal 
IMRT score obtained using the IO-TLP method and that obtained using the 
IO-SC method (see main text for description of these methods). 
}
\end{figure}

The feasibility of using this approximation for a BAO is demonstrated in 
Fig. \ref{LookupPinnacle} where we compare the optimal scores obtained 
using the IO-SC method with the corresponding scores obtained using 
the IO-TLP method. This plot has been generated using 100 randomly selected
sets of 3 beams that were used to plan a typical prostate case. 
Each score shown in Fig. \ref{LookupPinnacle} is the square root of the  
optimal value (according to the respective optimization
method) of the objective function which is the sum of contributions from all
DV constraints applied to the VOIs in this case (Wu and Mohan 2000). 
It can be seen that the IO-TLP scores are well correlated with the IO-SC
scores: ``good'' or ``bad'' sets of beam angles according to the IO-SC method 
are predicted similarly by the IO-TLP method. In general the IO-TLP scores are
lower than the IO-SC scores. This is due to the fact that less scatter
components are included in the IO-TLP method than in the IO-SC method.
Dose ``leakage'' outside the target, due to scatter, is slightly 
underestimated by the IO-TLP method and this gives rise to the undercounting
of the score and the finite variance of the correlation. It should be noted
that the systematic score undercounting in the IO-TLP method is practically
irrelevant to a BAO since only the relative {\it ranking} of different sets of beams
is needed. The presence of a finite variance in the correlation 
indeed introduces some limitation to the predicting power of the IO-TLP method.
It is the purpose of this paper, however, to show that even with 
this limitation the IO-TLP method can still be used as an effective, and
efficient, IMRT workhorse in a BAO system.

It can be seen in figure \ref{Flowchart} that our BAO algorithm
simply serves as an additional ``outer loop'' to a normal IMRT 
optimization (Wu and Mohan 2000). The beam-profile optimization
component of the BAO, however, uses a different dose-calculation
engine than the one that is used in normal IMRT optimization. 
The IO-TLP method is used as the beam-profile optimizer during
a BAO, because of its speed, while a normal IMRT optimization 
(for a given set of fixed beam angles) should be performed using 
the IO-SC method, because of its accuracy.

\begin{figure}
\begin{center}
\end{center}
\vskip -0.4in
\caption{\label{SAScore}Typical progression of the accepted score in a BAO
using an FSA algorithm as a function of the iteration number. 
The initial score is set to $T_0 = 1000$ in this run.
This is more than enough to allow all possible SA moves. As the optimization
proceeds, the temperature is lowered with a $T_0/(1 + i)$ schedule. As 
the temperature is lowered, SA practically turns into a local search algorithm 
although in FSA any point in the search space always has a finite probability to
be sampled at any temperature.}
\end{figure} 

It should be noted that although the IO-TLP method is used
during the BAO, a final beam-profile optimization using the IO-SC
method is performed on the optimal set of beam angles output 
by the BAO. This is a part of the finalization step shown in the
last box in figure \ref{Flowchart}. This final optimization 
of the BAO plan facilitates its comparison with a human-prepared 
plan, which is also optimized using the IO-SC method. 
In the following section, {\it all} of the displayed results  
have been produced using a IO-SC-based optimization.

\begin{figure}
\begin{center}
\end{center}
\vskip -0.4in
\caption{\label{3BeamsProstate}DVH comparison for coplanar beams. 
(a) Significant improvement can be obtained
for a 3-beam prostate plan. The DVHs for 3 equispaced beams are shown by
the dotted and dash-dotted lines, while those for the optimal BAO beams
are shown with solid lines. BAO provides clear improvement for all of
the DVHs. (b) Comparison of different sets of equispaced beams. In this
case, improved sparing of the rectum results in worse sparing of the 
bladder, or vice versa. The DVHs have been normalized to 72 Gy at 
90\% of target.} 
\end{figure}

\begin{table}
\caption{\label{SystemComparison}Comparison of the prostate and the head-and-neck 
cases that have been used in this work to test our BAO system.}
\begin{indented}
\item[]\begin{tabular}{@{}lcc}
\br
Parameter & Prostate & Head-and-neck\\
\mr
Number of CT slices   & 26         & 129          \\
Dose resolution       & 0.2 $\times$ 0.2 $\times$ 0.2 cm$^3$ & 0.2 $\times$ 0.2 $\times$ 0.2 cm$^3$ \\
Dose matrix dimension & (51,87,62) & (175,91,119) \\
Size of bitmap vector & 98,985     & 388,238     \\
Intensity matrix size (per beam) & $\sim (20 \times 15)$   & $\sim (85 \times 45)$    \\
Intensity resolution  & 0.3 $\times$ 0.5 cm$^2$ & 0.3 $\times$ 0.5 cm$^2$ \\
Target volume (cc)    & 91 (Prostate) & 62 (GTV), 175 (CTV) \\  
OAR volume (cc)       & 252 (Bladder) & 24 (Cord), 843 (Nodes) \\
                      & 78 (Rectum)   & 32, 17 (Right, Left Parotid) \\
\br
\end{tabular}
\end{indented}
\end{table}

\subsection{Materials}
\label{MaterialsSection}

Two cases have been used in this work to test the practicality 
of our BAO system: a prostate case and a head-and-neck case. 
Some of the system parameters of these two cases are displayed in 
table \ref{SystemComparison}. The prostate case represents the cases 
with relatively small size and simple arrangement of the target
and the OARs. The head-and-neck (HN) case represents the other end of the
spectrum namely those with relatively large size and with multiple 
targets that are located close to sensitive OARs. For the prostate case
we compare the BAO results for 3, 5, and 7 coplanar beams with that
of the corresponding number of equispaced coplanar beams. We also 
examine the dependence of the improvements that can be obtained from 
BAO on the criteria used to guide the optimization. This is done by 
studying the results of two different sets of criteria: 
a ``soft'' and a ``hard'' set. For the HN case we compare the BAO
results for 7 and 9 {\it non}-coplanar beams with that of 
9 equispaced coplanar beams.

The computational
time required to perform a BAO on a given case generally scales with 
the system size. This is determined principally by the 
product of the total number of beamlet intensities
that need to be optimized and the total number of voxels for 
which dose needs to be computed during the optimization. 
Note that table \ref{SystemComparison} lists the {\it typical} size
of the intensity matrix (which is the size of the smallest rectangle
that encloses the target in the BEV of the beam) since the specific 
size generally varies with the orientation of the beam. 
Furthermore, the number of beamlet intensities that are optimized
is generally less than the size of the intensity matrix because the 
beamlets that do not pass through any voxel in the target can be
completely turned off. Although increasing the number of optimization 
criteria is expected to increase the
total optimization time, the increase is minimal because once the dose
is calculated the value of the OFs can be computed very quickly. 
The size of the bitmap vector for the HN case in table 
\ref{SystemComparison} is about 4 times that of the prostate
case and the size of its intensity matrix is about $4 \times 3$ 
times larger. Thus, for the same number of beams, the total IMRT 
optimization time for the HN case is expected to be about 
48 times that of the prostate case. 
  
\section{Results}

\begin{figure}
\begin{center}
\end{center}
\vskip -0.4in
\caption{\label{5And7BeamsProstate}DVHs produced by BAO for
(a) 5; and (b) 7 coplanar beams compared to those produced by the 
corresponding number of equispaced beams for a prostate case 
(the same case as in figure \ref{3BeamsProstate}). The non-equispaced
beams selected by the BAO are better able to satisfy the 
specified DV criteria (table \ref{ProstateCriteria}). The
DVHs have been normalized to 72 Gy at 90\% of target.}
\end{figure} 

\subsection{Prostate case}

\label{prostatecase}

Figures \ref{3BeamsProstate} and \ref{5And7BeamsProstate}
display the improvements that can be obtained from BAO for
a prostate case. Each of the results shown in these figures 
has been obtained from a 1000-iteration FSA run and was completed in 
about 1--2 hours on a 400 MHz workstation (table \ref{CPUTime}). 
The initial SA temperature
$T_0$ was set to an arbitrary large number ($T_0 = 1000$) 
which allowed practically all random moves generated
by the SA sampling distribution to be accepted. This is an 
important practical requirement of the SA algorithm: The initial 
temperature has to be high enough to allow random sampling of the
search space during the initial iterations. Subsequently, the 
temperature was lowered using a $T_0/(1 + i)$ schedule. Figure 
\ref{SAScore} displays a typical progression of the 
score with increasing iteration number. The optimization is 
stopped after 1000 iterations (final temperature is 1) 
because practically the temperature is already too low to allow
a further significant change in the score to occur. Note that, based
on our experience as shown in figure \ref{SAScore}, the optimization
can generally be terminated after a few hundred iterations without
incurring a significant cost in the final score. 

\begin{table}
\caption{\label{CPUTime}Comparison of user CPU times for BAO with 1000 
sets of beams. Note that additional time is also needed to calculate 
the kernel for the beams in the search space: about 15 minutes for 
180 beams in the prostate case and 2 hours for 180 for 216 beams in 
the HN case.}
\begin{indented}
\item[]\begin{tabular}{@{}lcc}
\br
Case&CPU Speed &User CPU Time\\
\mr
3-Beam Prostate &400 MHz& 1 h 5 min \\
5-Beam Prostate &400 MHz& 1 h 40 min \\
7-Beam Prostate &400 MHz& 2 h 20 min \\
7-Beam HN       &750 MHz& 13 h 15 min \\
9-Beam HN       &750 MHz& 15 h 0 min\\
\br
\end{tabular}
\end{indented}
\end{table}

\begin{table}
\caption{\label{ProstateCriteria}DV criteria used for optimizing the prostate plan.}
\begin{indented}
\item[]\begin{tabular}{@{}lcccc}
\br
VOI&Type&Dose (cGy)&Volume (\%)&Penalty\\
\mr
Target & lower & 7200 & 90 &10\\
 & upper & 7900 & 5  &3\\
Bladder & upper & 6000 & 1 & 5\\
 & upper & 4000 & 5 & 3\\
Rectum & upper & 6000 & 5 & 5\\
 & upper & 4000 & 8 & 5\\
\br
\end{tabular}
\end{indented}
\end{table}

The optimal beams for this prostate case have been chosen from 
180 coplanar 18-MV photon beams (at zero couch angle)
which are distributed uniformly over the $360^\circ$ range for
the gantry angle with a $2^\circ$ interval. To facilitate the
comparison, each plan shown has been normalized to 72 Gy at 90\%
of the target volume. The DV-criteria that were used for the
IMRT optimization are listed in table \ref{ProstateCriteria}.

Figure \ref{3BeamsProstate}a compares the DVHs of a set of 3 
equispaced beams and those of a set of non-equispaced
beams selected by the BAO. It can be seen that the 
BAO beams provide significant improvements over the 
equispaced beams. The target receives a more uniform coverage
while both of the OARs, the bladder and rectum, receive
significantly less dose. Since in a 3-beam equispaced set
each beam is separated from its neighbors by a relatively
large $120^\circ$ interval, it is important to examine
if better results can be obtained from equispaced beams
by uniformly rotating the complete set. This is done in
figure \ref{3BeamsProstate}b where we compare 4 different
sets of equispaced beams, each equispaced configuration
is rotated by $30^\circ$ from the previous one. 
While their target coverages are
similar, they spare the OARs rather differently. It is interesting 
to observe that the set that provides better protection for 
the bladder does so at the expense of the sparing of the 
rectum, or vice versa. In addition, all of these plans 
are still inferior to that of the BAO beams shown in
figure \ref{3BeamsProstate}a. These comparisons illustrate
the importance of non-equispaced beam-angle selection in 
IMRT.  

DVH comparisons between equispaced and BAO plans for the 
same prostate case using 5 and 7 beams are presented in figures 
\ref{5And7BeamsProstate}a and \ref{5And7BeamsProstate}b
respectively. These plans have been optimized using the same
set of DV criteria as that used for the 3-beam plans in figure 
\ref{3BeamsProstate}. Although it is expected that the importance
of BAO will diminish as more beams are used (Stein \etal 1997), 
these cases show that BAO can still provide rather significant 
improvements over the equispaced beams. We have also performed
similar comparisons as in figure \ref{3BeamsProstate}b between
different sets of 5 and 7 equispaced beams. As is expected,
the difference among different 5- and 7-beam sets is smaller than 
that for the 3-beam set and it decreases as we increase the number
of beams from 5 to 7. 

\begin{table}
\caption{\label{Noncoplanar}BAO gantry and couch angles (in degrees) for noncoplanar HN plans.}
\begin{indented}
\item[]\begin{tabular}{@{}lccccccccc}
\br
Beam Number& 1 & 2 & 3 & 4 & 5 & 6 & 7 & 8 & 9\\
\mr
Gantry (9 Beams) & 5 & 100 & 140 & 205 & 240 & 245 & 290 & 300 & 345 \\
Couch (9 Beams) & 0 & 45 & 45 & 45 & 45 & 315 & 45 & 45 & 0 \\
Gantry (7 Beams) & 20 & 105 & 115 & 245 & 320 & 325 & 340 & - & - \\
Couch (7 Beams) & 315 & 0 & 45 & 0 & 45 & 315 & 45 & - & - \\
\br
\end{tabular}
\end{indented}
\end{table}

\begin{table}
\caption{\label{HNCriteria}DV criteria used for optimizing the head-and-neck plan.}
\begin{indented}
\item[]\begin{tabular}{@{}lcccc}
\br
VOI&Type&Dose (cGy)&Volume (\%)&Penalty\\
\mr
GTV           & lower & 7300 & 99 & 50\\
              & upper & 7700 & 0.1  & 30\\
CTV           & lower & 6200 & 96 & 50\\
              & upper & 6900 & 40 & 20\\
Nodes         & lower & 5600 & 95 & 50\\
              & upper & 5800 & 15 & 20\\
Cord          & upper & 3000 & 0.1 & 50\\
Brainstem     & upper & 3000 & 0.1 & 50\\
Left Parotid  & upper & 2500 & 30 & 20\\
              & upper & 5000 & 1 & 20\\
Right Parotid & upper & 1000 & 32 & 50\\
              & upper & 3000 & 10 & 50\\
\br
\end{tabular}
\end{indented}
\end{table}

\begin{table}
\caption{\label{HNPrescription}Prescription for the head-and-neck plan ($R_x$ = 6810 cGy).}
\begin{indented}
\item[]\begin{tabular}{@{}lc}
\br
DV& Prescription (cGy)\\
\mr
D$_{98}$(GTV) & $\geq$ 6810 \\
D$_2$(GTV) & $\leq$ 7491 \\
D$_{95}$(CTV) & $\geq$ 6000 \\
D$_{90}$(Nodes) & $\geq$ 5400 \\
D$_1$(Cord) & $<$ 4500 \\
D$_1$(Brainstem) & $<$ 5500 \\
\br
\end{tabular}
\end{indented}
\end{table}

\begin{figure}
\begin{center}
\end{center}
\vskip -0.4in
\caption{\label{9BeamsHN}Comparison between DVHs produced by 9
equispaced coplanar beams and those produced by 9 non-coplanar
beams selected by BAO for a head-and-neck case. The DVHs have
been normalized to 60 Gy at 95\% of the CTV volume. Solid lines
are for the 9 non-coplanar beams selected by the BAO (table \ref{Noncoplanar})
and dashed lines are for the 9 equispaced coplanar beams.} 
\end{figure} 

\begin{figure}
\begin{center}
\end{center}
\vskip -0.4in
\caption{\label{ContoursHN}HN isodose contours for coplanar equispaced
plan (top) compared with those of non-coplanar BAO plan (bottom). 
Nine beams are used in each plan. BAO plan provides 
significantly better sparing of the parotid glands compared
to that of the equispaced beams.}
\end{figure}

\begin{figure}
\begin{center}
\end{center}
\vskip -0.4in
\caption{\label{7BeamsHN}Comparison between DVHs produced by 9
equispaced coplanar beams and those produced by 7 non-coplanar
beams selected by BAO for a head-and-neck case. The DVHs have
been normalized to 60 Gy at 95\% of the CTV volume. Solid 
lines are for the 7 non-coplanar beams selected by the BAO 
(table \ref{Noncoplanar}) and dashed lines are for the 
9 equispaced coplanar beams.}
\end{figure}

\subsection{Head-and-neck case}

Due to its anatomical position, head-and-neck tumors are
good candidates for the application of non-coplanar IMRT.
Limited experience with non-coplanar HN plans and the fact that
selection of non-coplanar beam directions for IMRT is a highly
non-intuitive process also make global search methods like the FSA,
which is implemented in our BAO system, particularly attractive.

Figures \ref{9BeamsHN} and \ref{7BeamsHN} compare the DVHs
for a HN case of 9 and 7 non-coplanar beams, respectively, 
with those of a benchmark 9 equispaced coplanar beams. The DV 
criteria used to optimize these HN plans are listed in table 
\ref{HNCriteria} while the gantry and couch angles selected
by our BAO algorithm for these sets are tabulated in table 
\ref{Noncoplanar}. Each of the BAO results for this HN case
has been obtained from a 1000-iteration FSA runs. The optimal
beams have been chosen from 216 possible directions: the  
search pool for this case consists of 3 possible couch angles
($0^\circ$, $45^\circ$, and $315^\circ$) with 72 possible
gantry angles distributed uniformly with $5^\circ$ interval
for each couch angle. In principle, finer resolution for the
couch angle may be used. However, this will also increase the
memory requirement since our BAO system, as it is currently 
implemented, requires the kernel for each beam to be available
for quick access in the system memory. For this HN case, the 
size of the kernel is about 3 MB for each beam and therefore
it requires about 650 MB of memory to store all the kernels.
We have used a 750-MHz Sun workstation to perform the BAO 
for this HN case. The 1000-iteration BAO run for the 9 
non-coplanar beams required 17 hours to complete (this includes
about 2 hours of computing time required for kernel extraction) while the
1000-iteration run for the 7 non-coplanar beams required 13 hours
(kernels were read from disk for this case). 
The relatively long optimization time for this HN
case, compared to the corresponding time for the prostate case, 
is due to the substantially larger size of the system.
A side-by-side comparison of these two systems is presented
in table \ref{SystemComparison}. Note that, as pointed
out previously in the discussion regarding figure \ref{SAScore}, 
the number of iterations used can be substantially reduced 
without significant reduction in the optimal score. This has 
the potential of shortening the total optimization time by a factor
of two. Further reduction in time can be obtained from using a more selective
search, e.g. not allowing the beams to be too close to each other,
inclusion of prior knowledge, and utilization of parallel processors.
These possible avenues for improvements are currently being investigated.

Figure \ref{9BeamsHN} compares the DVHs for a set of 9 
non-coplanar BAO beams with those of a 9-beam equispaced coplanar set.
The BAO set provides both better dose uniformity at the GTV and CTV and 
better protection of the parotid glands. This better sparing
of the parotid glands can also be seen in the isodose contours shown
in figure \ref{ContoursHN}. Figure \ref{7BeamsHN}
shows that a plan of similar quality can also be obtained 
using 7 non-coplanar beams, although in this case the quality
of target coverage is lower than that of the 9 non-coplanar beams.  
The dose prescriptions for these plans are shown in table 
\ref{HNPrescription}. The plans have been normalized to best satisfy
all of these prescriptions. These have been achieved by normalizing
the DVHs to 60 Gy at 95\% of the CTV volume. 

\section{Discussion}

\begin{table}
\caption{\label{SoftCriteriaTable}``Soft'' DV criteria used for optimizing the prostate plans
shown in figure \ref{SoftCriteriaPlot}. The criteria for the target are the same as in 
table \ref{ProstateCriteria} while those for the bladder are ``softer''. Only one of the 
criteria for the rectum in table \ref{ProstateCriteria} is used here.}
\begin{indented}
\item[]\begin{tabular}{@{}lcccc}
\br
VOI&Type&Dose (cGy)&Volume (\%)&Penalty\\
\mr
Target & upper & 7200 & 90 &10\\
 & lower & 7900 & 5  &3\\
Bladder & lower & 6000 & 3 & 5\\
 & lower & 4000 & 10 & 1\\
Rectum & lower & 6000 & 5 & 5\\
\br
\end{tabular}
\end{indented}
\end{table}

\begin{figure}
\begin{center}
\end{center}
\vskip -0.4in
\caption{\label{SoftCriteriaPlot} Two sets of 3 beam angles that were selected 
by the BAO as ``good'' sets when ``soft'' DV criteria (table \ref{SoftCriteriaTable}) 
were used for the IMRT optimization. These illustrate the importance of good 
selection of DV criteria in BAO. The DVHs have been normalized to 72 Gy at 
90\% of the target.}
\end{figure}

\subsection{Dependence on criteria}

A noteworthy feature of a DV-based optimization is that further improvement will 
no longer be attempted by the optimizer once the criteria are satisfied. In other
words, once the contribution of a DV criterion to the OF reaches zero, it 
becomes irrelevant to the optimizer. Consider, e.g., a single criterion which 
is assigned to the rectum: $< 6000$ cGy at 5\% of its volume. Once this 
criterion is satisfied, e.g., when $D_5$(rectum) has been lowered to 5900 cGy, 
the optimizer will no longer attempts to improve the 
dose to the rectum, even though it is still possible to reduce its dose further.  

We have found that the improvement on plan quality that is output by our BAO 
system generally depends on the DV criteria that we assign to the VOIs.
Figure \ref{SoftCriteriaPlot} compares two of the top 5 plans generated 
by the BAO (for the prostate case presented in section 
\ref{prostatecase}) using several ``soft'' criteria listed in table \ref{SoftCriteriaTable},
with the corresponding plan for a set of 3 equispaced beams. These plans have 
been generated for the same prostate case as that used in figures 
\ref{3BeamsProstate} and \ref{5And7BeamsProstate}. The BAO sets are both
predicted, by the IO-TLP method which is used during the BAO, 
to have lower score compared to the equispaced beams (55 and 75 
for plans shown in figure \ref{SoftCriteriaPlot}a and \ref{SoftCriteriaPlot}b,
respectively, compared to 186 for the equispaced beams). However, upon 
reoptimization using the IO-SC method, the BAO sets produce higher 
score than the equispaced beams (186 and 212 for figure \ref{SoftCriteriaPlot}a
and \ref{SoftCriteriaPlot}b, respectively, compared to 162 for the
equispaced beams).

The large score increase encountered in going from the IO-TLP to the 
IO-SC method for the plan shown in figure \ref{SoftCriteriaPlot}a 
is largely due to the additional score accumulated by the DVH of the
rectum. In this case, the IO-TLP method minimized the score by creating
a steep gradient of the DVH curve just below the 6000 cGy dose. Note that even
though below 6000 cGy the BAO DVH for the rectum is {\it higher}, and 
therefore is worse, than the corresponding DVH of the equispaced beams, 
the contribution of this DVH to the OF can be {\it smaller} if its DVH 
at 6000 cGy is lower than the corresponding value for the equispaced beams.
This is specifically what happens in this case due to the steep gradient
of the BAO DVH curve. In going from the IO-TLP to the IO-SC method,
however, this steep gradient cannot be sustained because of the inclusion
of more scatter components in the IO-SC method. This ``dose leakage''
substantially increases the partial volume of the rectum that receives
dose greater than 6000 cGy and, consequently, produces a large 
increase in score. 

The BAO plan in figure \ref{SoftCriteriaPlot}b satisfies the criteria 
listed in table \ref{SoftCriteriaTable} as well as, or indeed better
than, that of the equispaced beams. In this case, the higher score
of the BAO plan is due to the contribution from the ``normal tissue'' 
(NT) region. The NT region encompasses generic normal tissue surrounding
the target that does not belong to any particular OAR (Wu and Mohan 2000). 
For the prostate case that we use in this work, the NT is a 1 cm extension
of the target that is not part of any OAR and it is assigned an upper constraint 
of 6000 cGy at 5\% of its volume. The ``dose leakage'' from 
the target to its surrounding NT, for the case shown in figure 
\ref{SoftCriteriaPlot}b is substantially larger than the allowance set by the
assigned DV criterion and it therefore introduces a significant 
contribution to the OF on going from the IO-TLP to the IO-SC method. 
This is due to the small angular separation of 
two of the 3 beams in the BAO set ($238^\circ$ and $254^\circ$) that
provides an elongated dose distribution along the direction of these beams.

In contrast to this relatively large uncertainty in the prediction of the 
IO-TLP method for cases with ``soft'' DV criteria, in our experience
we found that its prediction is much more reliable for cases with 
more demanding criteria, such as listed in table \ref{ProstateCriteria}
and displayed in figures \ref{3BeamsProstate} and \ref{5And7BeamsProstate}.
Although we only present this dependence on criteria using a prostate-case
example in this paper, similar behavior has also been observed for the HN case. 
One possible reason that may be given for this observation is that
the degeneracy of the solution, i.e. the multiminima of the OF, is 
larger for ``soft'' criteria (Wu and Mohan 2002). A careful study, 
however, is necessary in order to provide a definitive answer to this 
question. From the point of view of clinical applications, the fact 
that BAO is more useful for ``hard'' cases is not expected to 
eliminate its practical relevance. It is well known that ``soft'' 
cases can be planned satisfactorily with equispaced beams or, indeed, 
class solutions. On the other hand, for ``hard'' cases, e.g. those 
of re-irradiation or of complicated geometry, class solutions are 
non-existent (almost by definition) and equispaced beams may not 
provide the optimal solution. For these cases the fast BAO described
in this paper provides an attractive alternative to a manual 
trial-and-error approach.  

\subsection{Further speed up?}

Several schemes have been proposed recently to rank beam directions according to
their {\it individual} potential for creating a desirable IMRT distribution
such as the single-beam cost function (Oldham \etal 1998), beam's-eye-view 
dosimetrics (BEVD) (Pugachev and Xing 2001), and the figure-of-merit
parameter (Chuang \etal 2003). These schemes have also been studied
as a possible way to intelligently accelerate a BAO. If a particular beam direction
can be reliably predicted to always give rise to high values of objective function,
independent of the directions of the other beams in the set it is a member of, it 
can be quickly removed from the search space. It is perhaps important
to point out here, in connection with the OF of the 
optimization, that these schemes are expected to have less significance
in a DV-based optimization than in a dose-based optimization for which
they have been proposed and studied. Individual ranking of beam directions
is useful in a dose-based optimization where each VOI is assigned 
a specific prescribed dose. In contrast, in a DV-based optimization
each VOI is effectively assigned a specific {\it range} of acceptable 
doses. In this case, several different distributions of the total dose 
may have equally good value of the DV OF (in other words,
the OF possesses multiple local minima and degenerate solutions). 
It is therefore less meaningful to attach 
a specific geometric ``shape'', and therefore potentially ``good'' or ``bad'' 
beam directions for creating this shape, to the optimal distribution.

No a priori knowledge has been used to guide the selection of 
beam angles in this work. The use of reliable prior knowledge
is expected to further improve the speed and potential of our BAO
system. However, the complexity of the BAO problem requires 
judicious selection, and therefore careful systematic study, of 
what ``knowledge'' is best to be used for each given case.
As has been discussed in this paper, this systematic study is especially 
more pertinent to the DV-based IMRT optimization that we use in 
our BAO system, because of the presence of the multiple local minima
of the OF, than to the dose-based optimization
that is commonly used in other BAO systems.
In our BAO system, the beam angles are selected from the preset
values in the search space. Some prior knowledge can be 
incorporated in the initial selection of beams that are 
to be included in the search space, e.g., more and less beams 
can be positioned in favorable and unfavorable directions, respectively. 
Alternatively, the beam angles selected during the optimization 
can be restricted to have a minimum separation from each other.
Prior knowledge can also be used to reduce the number of beams
in the search pool. Although large search pools are used in this study,
180 and 216 beams for prostate and HN case respectively, 
in a clinical setting one may not need to use such large numbers.
Using a reasonable spacing of 10$^\circ$ between adjacent 
beams in a coplanar pool, e.g., allows us to reduce the size of the
search pool to 36. This will significantly decrease the number of 
possible {\it combinations} of beam angles to be sampled during
the BAO. In the current version of our BAO system, each beam in the trial
set is given a uniform initial profile prior to the beam-profile optimization.
Instead of this uniform intensity distribution, the initial beam profile can be set
to a {\it non}-uniform one, chosen based on some prior knowledge. 
This has the potential of reducing the number of iterations needed
for the beam-profile optimization and, consequently, the total
BAO time. 

\section{Conclusion}


In this paper we have demonstrated that a combination of a 
fast table-lookup dose calculation method and the FSA algorithm can 
be used in a practical BAO system. This method has the advantage of 
being able to produce clinically significant improvement within
a reasonable time. We showed, e.g., that the beam angles of a 7-beam
prostate plan can be optimized within about 2 hours on a moderate
400 MHz workstation. A DV-based OF is used to guide the IMRT 
optimization in our BAO system. This provides a direct link to
our clinical IMRT system and facilitates the comparison between
the BAO plans and those produced by human planners. Since the 
beam-angle optimizer in our system simply wraps around the 
IMRT beam-profile optimizer, other OFs that are
available to our IMRT optimizer, such as EUD, can also be used 
for BAO in our system. Our BAO system can also search for
the optimal set over non-coplanar beams. This feature may be
particularly helpful to planners because of the insufficient experience
with non-coplanar beams in treatment planning.    

\bigskip

\noindent {\bf Acknowledgements}

\bigskip

\noindent This work is supported in part by grant CA74043 from the National
Cancer Institute and a research grant from Philips Radiation Oncology Systems.


\section*{References}

\begin{harvard}

\item[] Bortfeld T and Schlegel W 1993 Optimization of beam orientations in 
radiation therapy: some theoretical considerations \PMB {\bf 38} 291--304

\item[] Bortfeld T, Stein J and Preiser K 1997 Clinically relevant intensity
modulation optimization using physical criteria {\it XII International
Conference on the Use of Computers in Radiation Therapy} (Medical Physics
Publishing, Madison, WI) 1--4

\item[] Braunstein M and Levine R Y 2000 Optimum beam configurations in tomographic
intensity modulated radiation therapy \PMB {\bf 45} 305--328

\item[] Chuang K S, Chen T J, Kuo S C, Jan M L, Hwang I M, Chen S, Lin Y C and Wu J 2003
Determination of beam intensity in a single step for IMRT inverse planning 
\PMB {\bf 48} 293--306 

\item[] Crooks S M and Xing L 2001 Linear algebraic methods applied to intensity 
modulated radiation therapy \PMB {\bf 46} 2587--2606

\item[] Das S, Cullip T, Tracton G, Chang S, Marks L, Anscher M and Rosenman J 2003
Beam orientation selection for intensity-modulated radiation therapy based on
target equivalent uniform dose maximization {\it Int. J. Radiat. Oncol. Biol. Phys.}
{\bf 55} 215--224

\item[] Geman S and Geman D 1984 Stochastic relaxation, Gibbs distributions, and 
Bayesian restoration of images {\it IEEE Trans. Patt. Anal. Mach. Intell.}
{\bf 6} 721--741

\item[] Jackson A, Wang X H and Mohan R 1994 Optimization of conformal treatment
planning and quadratic dose objectives {\it Med. Phys.} {\bf 21} 1006

\item[] Kirkpatrick S, Gelatt C D and Vecchi M P 1983 Optimization by simulated 
annealing {\it Science} {\bf 220} 671--680

\item[] Llacer J, Deasy J O, Bortfeld T R, Solberg T D and Promberger C 2003
Absence of multiple local minima effects in intensity modulated optimization
with dose-volume constraints \PMB {\bf 48} 183--210

\item[] Mageras G S and Mohan R 1993 Application of fast simulated
annealing to optimization of conformal radiation treatments {\it Med. Phys.} {\bf 20}
639--647

\item[] Metropolis N, Rosenbluth A W, Rosenbluth M N, Teller A H and
Teller E 1958 Equation of state calculation by fast computing machines 
{\it J. Chem. Phys.} {\bf 21} 1087--1092

\item[] Oldham M, Khoo V, Rowbottom C G, Bedford J and Webb S 1998 
A case study comparing the relative benefit of optimising beam-weights, 
wedge-angles, beam orientations and tomotherapy in stereotactic radiotherapy
of the brain \PMB {\bf 43} 2123--46

\item[] Pugachev A B, Boyer A L, and Xing L 2000 Beam orientation optimization in 
intensity-modulated radiation treatment planning {\it Med. Phys} {\bf 27}
1238--1245 

\item[] Pugachev A, Li J G, Boyer A L, Hancock S L, Le Q T, Donaldson S S 
and Xing L 2001 Role of beam orientation optimization in intensity modulated
radiation therapy \PMB {\bf 50} 551--560

\item[] Pugachev A and Xing L 2001 Pseudo beam's-eye-view as applied to 
beam orientation selection in intensity-modulated radiation therapy
{\it Int. J. Radiat. Oncol. Biol. Phys.} {\bf 51} 1361--1370

\item[] \dash 2002 Incorporating prior knowledge into beam
orientation optimization in IMRT {\it Int. J. Radiat. Oncol. Biol. Phys.}
{\bf 54} 1565--1574

\item[] Rowbottom C G, Webb S and Oldham M 1998 Improvements in prostate
radiotherapy from the customization of beam directions {\it Med. Phys}
{\bf 25} 1171--1179

\item[] \dash 1999a Beam-orientation customization using an artificial
neural network \PMB {\bf 44} 2251--2262

\item[] Rowbottom C G, Oldham M and Webb S 1999b Constrained customization
of non-coplanar beam orientations in radiotherapy of brain tumors
\PMB {\bf 44} 383--399

\item[] Spirou S V and Chui C S 1998 A gradient inverse planning algorithm
with dose-volume constraints {\it Med. Phys.} {\bf 25} 321--333

\item[] Stein J, Mohan R, Wang X H, Bortfeld T, Wu Q, Preiser K, Ling C C
and Schlegel W 1997 Number and orientations of beams in intensity-modulated
radiation treatments {\it Med. Phys.} {\bf 24} 149--160

\item[] Szu H and Hartley R 1987 Fast simulated annealing {\it Phys. Lett. A}
{\bf 122} 157--162

\item[] Webb S 1989 Optimisation of conformal radiotherapy dose distributions
by simulated annealing \PMB {\bf 34} 1349--1370

\item[] \dash 1991 Optimization by simulated annealing of three-dimensional
conformal treatment planning for radiation fields defined by a multileaf collimator
\PMB {\bf 36} 1201--1226

\item[] Wu Q and Mohan R 2000 Algorithms and functionality of an intensity 
modulated radiotherapy optimization system {\it Med. Phys.} {\bf 27} 701--711

\item[] \dash 2002 Multiple local minima in IMRT optimization based on dose-volume
criteria {\it Med. Phys.} {\bf 29} 1514--1527

\item[] Wu Q, Djajaputra D, Lauterbach M, Wu Y and Mohan R 2003
A fast dose calculation method based on table lookup for IMRT optimization
\PMB {\bf 48}, N159--166

\item[] Xiang Y and Gong X G 2000 Efficiency of generalized simulated 
annealing {\it Phys. Rev. E} {\bf 62} 4473--4476

\item[] Xing L and Chen G T Y 1996 Iterative methods for inverse treatment
planning \PMB {\bf 41} 2107--2123

\end{harvard}

\end{document}